\documentclass[
tightenlines,
twocolumn,
a4paper,
nofootinbib,
superscriptaddress,
floatfix,
preprintnumbers,
flushbottom,
aps
]{revtex4-1}

\usepackage{url}
\usepackage[%
        hyperindex,breaklinks,
	pdfstartview=FitH,
  pdfpagemode=UseNone
	]{hyperref}

\usepackage[
   centertags, 
   sumlimits,  
   intlimits,  
   namelimits, 
]{amsmath} %
\usepackage{amssymb}

\makeatletter
\def\tagform@#1{\maketag@@@{\ignorespaces#1\unskip\@@italiccorr}}
\let\orgtheequation\theequation
\def\theequation{(\orgtheequation)}
\makeatother

\usepackage[all]{hypcap} 
\renewcommand{\BibitemShut}[1]{}

\usepackage[english]{babel} 

\usepackage[%
]{graphicx}

\begin{document}
\title{Momentum imbalance of D mesons in ultrarelativistic heavy-ion collisions at the CERN Large Hadron Collider}

\author{Jan Uphoff}

\author{Florian Senzel}
\affiliation{Institut f\"ur Theoretische Physik, Goethe-Universit\"at Frankfurt, Max-von-Laue-Str.\ 1, 
D-60438 Frankfurt am Main, Germany}

\author{Zhe Xu}
\affiliation{Department of Physics, Tsinghua University, Beijing 100084, China}
\affiliation{Collaborative Innovation Center of Quantum Matter, Beijing, China}

\author{Carsten Greiner}
\affiliation{Institut f\"ur Theoretische Physik, Goethe-Universit\"at Frankfurt, Max-von-Laue-Str.\ 1, 
D-60438 Frankfurt am Main, Germany}

\date{\today}

\begin{abstract}
As a new observable for heavy flavor correlations the momentum imbalance $A_D$ of $D$~mesons is proposed. It is defined analogously to the jet momentum imbalance $A_J$ of fully reconstructed jets.
However, since $D$~mesons are flavor tagged particles, no jet reconstruction is necessary.
$A_D$ quantifies the influence of the medium created in heavy-ion collisions on correlated charm pairs. 
We present results with the partonic transport model \emph{Boltzmann Approach to MultiParton Scatterings} (BAMPS), which describes well the nuclear modification factor and elliptic flow of all heavy flavor particles at RHIC and LHC. The $A_D$ distribution in heavy-ion collisions at LHC is shifted to larger values of $A_D$ compared to proton-proton collisions. We argue that this shift is due to medium effects and can be explained partially by a path length imbalance of charm pairs and partially by momentum fluctuations in the initial charm pair distribution.
\end{abstract}


\maketitle

\section{Introduction}
Several experimental observations indicate that in ultra-relativistic heavy-ion collisions at the Large Hadron Collider (LHC) at CERN 
a new state of matter, i.e., the quark-gluon plasma (QGP), is produced \cite{Muller:2012zq}. 

High-energy particles in such heavy-ion collisions are created in initial hard parton scatterings  with a large momentum transfer and are therefore denoted as hard probes. While traversing the created QGP they interact with other partons and deposit lots of their energy in the medium. An observable for this energy loss is the nuclear modification factor $R_{AA}$ of single hadrons. It is defined as the yield in heavy-ion (A+A) collisions divided by the yield in proton-proton (p+p) collisions scaled with the number of binary collisions,
\begin{align}
\label{raa_def}
  R_{AA}=\frac{{\rm d}^{2}N_{\text{AA}}/{\rm d}p_{T}{\rm d}y}{N_{\rm bin} \, {\rm d}^{2}N_{\text{pp}}/{\rm d}p_{T}{\rm d}y} \ .
\end{align}
Another observable that is accessible at the LHC is the transverse momentum imbalance $A_J$ of fully reconstructed jets,
\begin{align}
A_J = \frac{p_{T;1}^J - p_{T;2}^J}{p_{T;1}^J + p_{T;2}^J} \ ,
\end{align}
where $p_{T;1}^J$ ($p_{T;2}^J$) is the transverse momentum of the leading (subleading) jet with the highest (second highest) transverse momentum in the measured rapidity window.

Due to the presence of the QGP the distribution of the momentum imbalance of jets in heavy-ion collisions is shifted to larger values compared to p+p collisions \cite{Aad:2010bu,*Chatrchyan:2011sx,*Chatrchyan:2012nia}. The picture beyond this phenomenon is the following: two hard partons are produced in an initial hard scattering under a large angle (back-to-back in the transverse plane in leading order perturbative QCD). Different path lengths of these two partons lead to different amounts of energy loss, which in turn increases the momentum imbalance \cite{Qin:2010mn,*CasalderreySolana:2011rq,*He:2011pd,*Young:2011qx,*Renk:2012cx,*ColemanSmith:2012vr,*Ma:2013pha,Senzel:2013dta}. 
However, interpreting the experimental data by comparing to theoretical calculations is challenging since the results are very sensitive to the jet finding algorithm, detector effects, and the background subtraction scheme \cite{Senzel:2013dta}. Therefore, it is necessary to employ in the theoretical simulations the same techniques as in the experimental data analysis to draw any conclusions.

Another type of hard probes are heavy quarks, i.e., charm and bottom quarks, which are also exclusively produced in hard processes due to their large mass \cite{Uphoff:2010sh}. Furthermore, charm and anti-charm quarks (or bottom and anti-bottom quarks) are always produced in pairs. After their early production, they traverse the medium, lose energy, and participate in the collective flow. Because of flavor conservation in QCD, heavy quarks are tagged particles, transferring their flavor during hadronization to $D$ and $B$ mesons, which allows the identification of heavy flavor particles in measurements.

\section{D meson momentum imbalance}
In this paper we propose as a new observable the transverse momentum imbalance~$A_D$ of $D$~mesons, which is defined analogously to $A_J$ as
\begin{equation}
 A_D = \frac{p_{T;1}^D - p_{T;2}^D}{p_{T;1}^D + p_{T;2}^D} \ ,
\end{equation}
where $p_{T;1}^D$ ($p_{T;2}^D$) is the transverse momentum of the leading (subleading) $D$~meson (or $\bar D$~meson) with the highest (second highest) transverse momentum in the considered rapidity window (in this paper $|y|<1$). Due to their unique property of being tagged particles, there is no need to perform a complicated jet reconstruction. However, such a jet reconstruction of charm tagged jets would reduce the imbalance that is already present in the vacuum without a QGP evolution and therefore might make the shift in the $A_D$ distribution from events with and without QGP more pronounced (cf.\ Figs.~\ref{fig:ad_10_4_23pi} and \ref{fig:ad_25_15_23pi}).

In principle, for this observable $D$~mesons should be measured on an event-by-event basis. Although $D$~mesons can only be statistically reconstructed, unwanted correlation from randomly associated background tracks at the $D$~meson mass could be distinguished from real $D$~mesons correlations by subtracting correlations from the side bands of the $D$~meson invariant mass distribution~\cite{Colamaria:2013una}. 
Due to small branching ratios of the reconstructed decay channels and due to a small signal to background ratio this is currently difficult, but future runs at LHC might offer enough statistics for that measurement. An alternative would be to trigger on the leading $D$ meson and measure associated heavy flavor electrons or high-energy charged hadrons on the away-side, which probably will be the associated $\bar D$ mesons if the trigger $p_T$ is high enough and the production mechanism is leading order\footnote{If charm is produced in next-to-leading order gluon splitting processes, the angle between both produced charm quarks might be small and therefore the leading hadron at the away side might not be the $D$~meson.}, even if it cannot be identified experimentally. 
Such $D$ meson and charged hadron correlation studies are currently performed in p+p and p+Pb collisions \cite{DeepaThomasfortheALICE:2013oga,Colamaria:2013una}.

As a note, $A_D$ is strongly related to heavy flavor correlations \cite{Zhu:2006er,*Zhu:2007ne,*Gossiaux:2009mk,*Younus:2013be,*Nahrgang:2013saa} of two charm quarks stemming from the same initial hard parton scattering. Experimentally, however, it is challenging to measure these correlations since one does not know which $D$ and $\bar D$~mesons are related. Therefore, one would have to perform tedious background subtractions. In contrast, for determining $A_D$ no such background subtraction of uncorrelated $D$~mesons is needed (however, a background subtraction of randomly associated background tracks at the $D$~meson mass might be necessary). Nevertheless, by modifying the trigger conditions for the leading and subleading $D$~mesons one has a handle on the fraction of $D$~meson pairs that passes the trigger and stem from a charm pair that is produced in the same hard parton scattering.

In the following we present our results for the $D$~meson momentum imbalance~$A_D$. After briefly introducing the partonic transport model \emph{Boltzmann Approach to MultiParton Scatterings} (BAMPS) and the considered heavy flavor processes, we show the $A_D$ distributions for different trigger conditions and initial heavy flavor distributions. Thereafter, we argue that the observed increased momentum imbalance~$A_D$ in A+A collisions compared to p+p collisions can be explained partially by different path lengths and partially by momentum fluctuations in the initial charm pair distribution.

\section{Parton cascade BAMPS}

The partonic transport model \emph{Boltzmann Approach to MultiParton Scatterings} (BAMPS) \cite{Xu:2004mz,*Xu:2007aa} describes the 3+1 dimensional evolution of the QGP phase by propagating all particles in space and time and carrying out their collisions according to the Boltzmann equation. The initial heavy quark distributions are either obtained with PYTHIA 6.4 \cite{Sjostrand:2006za} or the leading order mini-jet model. For the former we extract the heavy quarks from PYTHIA before hadronization sets in and put them into BAMPS. In the latter procedure we sample the initial heavy quarks according to leading order perturbative QCD (pQCD), which produces back-to-back heavy quark pairs. 

As a note, light partons interact inelastically with the original Gunion-Bertsch cross section \cite{Gunion:1981qs}. Recently, some issues with this approximation at forward and backward rapidity of the radiated gluon have been addressed \cite{Fochler:2013epa}. Updated BAMPS results on the light parton sector with the new improved Gunion-Bertsch cross section can be found in Ref.~\cite{Uphoff:2014cba}. However, we checked that the effect on the heavy quark $R_{AA}$ and $A_D$ due to the improvements on the light particle sector is only minor. 

For heavy quarks only elastic processes are considered in this paper. Their cross sections are calculated in leading order pQCD, explicitly taking the running coupling into account.
The divergent $t$ channel is regularized with a screening mass $\mu$ that is determined by matching elastic energy loss calculations with leading order pQCD cross sections to results from hard thermal loop calculations. The comparison of both results shows that the screening mass~$\mu$ is smaller than the usually employed Debye mass~$m_D$, more precisely, $\mu^2 = \kappa_t m_D^2$ with $\kappa_t = 1/(2e) \approx 0.2$. The hadronization of charm quarks to $D$ mesons is performed with Peterson fragmentation \cite{Peterson:1982ak} as described in Ref.~\cite{Uphoff:2011ad} with the commonly used parameter $\epsilon_c = 0.05$ for charm quarks. The details of the heavy flavor implementation in BAMPS can be found in Ref.~\cite{Uphoff:2011ad,Uphoff:2012gb}.

Since radiative heavy quark processes are currently being implemented in BAMPS and are not considered in this study, we mimic their influence by effectively increasing the elastic cross section by a factor $K=3.5$, which is tuned to  the  elliptic flow data of heavy flavor electrons at RHIC \cite{Uphoff:2012gb} and can also describe the electron nuclear modification factor at RHIC. Having fixed this parameter to the RHIC data, we find a good agreement with the experimentally measured nuclear modification factor of $D$~mesons, non-prompt~$J/\psi$ (from $B$~meson decays), and heavy flavor muons at the LHC energy of $\sqrt{s_{NN}}=2.76 \, {\rm TeV}$ \cite{Uphoff:2012gb}. Furthermore, we made predictions for the nuclear modification factor of heavy flavor electrons as well as the non-prompt~$J/\psi$, $D$~meson, electron, and muon elliptic flow at the LHC \cite{Uphoff:2012gb}, which agree well with recently released data \cite{delValle:2012qw,*Sakai:2013ata,*Abelev:2013lca}.

\section{Results}

\begin{figure}
\includegraphics[width=1.0\linewidth]{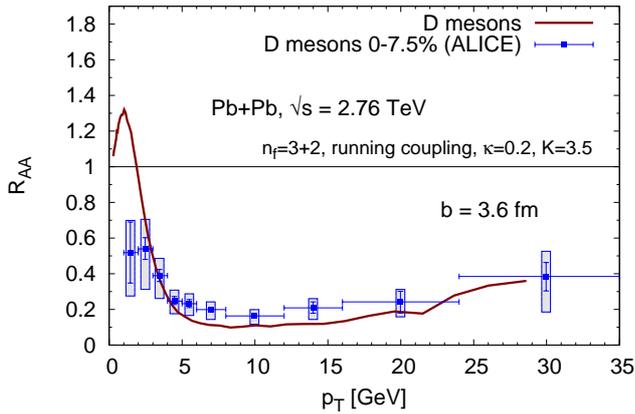}
\caption{(Color online) Nuclear modification factor $R_{AA}$ of $D$~mesons for Pb+Pb collisions at the LHC energy of $\sqrt{s_{NN}}=2.76 \, {\rm TeV}$ with impact parameter $b=3.6\,{\rm fm}$ together with data \cite{Grelli:2012yv}.
}
\label{fig:raa}
\end{figure}

In Fig.~\ref{fig:raa} we depict our predictions for the nuclear modification factor $R_{AA}$ of $D$~mesons for the most central events ($0-7.5$\,\% centrality class) of heavy-ion collisions at LHC, calculated with the same parameter as stated above and also as in Ref.~\cite{Uphoff:2012gb}.\footnote{As a note, at the time we made the prediction for the $0-7.5$\,\% centrality class the $D$~meson $R_{AA}$ for the centrality class $0-20$\,\% was already published \cite{ALICE:2012ab}, albeit with a much smaller transverse momentum outreach.}
We find a good agreement with the experimental data for intermediate and large transverse momenta. 

\begin{figure}
\includegraphics[width=1.0\linewidth]{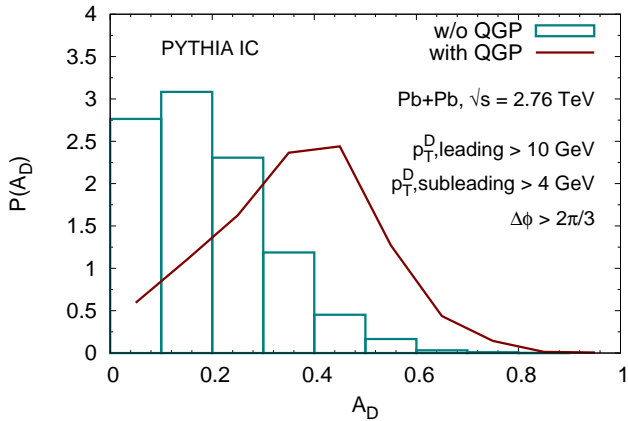}
\caption{(Color online) Distribution of the $D$~meson momentum imbalance~$A_D$ for Pb+Pb collisions at the LHC energy of $\sqrt{s_{NN}}=2.76 \, {\rm TeV}$ with impact parameter $b=3.6\,{\rm fm}$. The initial heavy quarks are obtained with PYTHIA. The trigger condition for the leading (subleading) $D$~meson is $p_{T;1}^D>10 \, {\rm GeV}$ ($p_{T;2}^D>4 \, {\rm GeV}$). In addition, the angle $\Delta\phi$ between these two $D$~mesons must be larger than $2\pi/3$.
}
\label{fig:ad_10_4_23pi}
\end{figure}

Figure~\ref{fig:ad_10_4_23pi} shows the distribution of the $D$~meson momentum imbalance~$A_D$ in heavy-ion collisions at LHC.
Even if no QGP were present, a sizable momentum imbalance is visible. This is due to initial and final state radiation at the production processes of charm quarks, which are simulated with PYTHIA \cite{Sjostrand:2006za}. However, if interactions of charm quarks with other partons of the QGP 
are also allowed, the curve is shifted to larger values of the momentum imbalance. Thus, the presence of the medium in heavy-ion collisions increases the momentum imbalance of $D$~mesons considerably.

In contrast to experiments, in which only the final $D$~mesons are measured, we have in BAMPS access to the full evolution history of each charm quark and corresponding $D$~meson. Therefore, we can check if the two $D$~mesons with the highest transverse momenta that pass the trigger conditions stem from two charm quarks that have been produced in the same initial hard scattering. This is only true for 15\,\% of the $D$~meson pairs depicted in Fig.~\ref{fig:ad_10_4_23pi} due to the abundance of charm quarks in heavy-ion collisions at LHC and the relatively low triggers. Thus, most of the $D$~meson pairs that pass the trigger are from two charm quarks that are not related.

Still, Fig.~\ref{fig:ad_10_4_23pi} reveals that the $D$~meson momentum imbalance is strongly modified by the QGP and that the subleading jet loses a sizable amount of energy. Therefore, it would be highly interesting to measure this momentum imbalance in heavy-ion collisions at LHC. 

However, from a theoretical point of view it would be even more enlightening if the two measured $D$~mesons with the highest momenta would stem from the same initial hard event. The most natural way to ensure this is to increase the trigger conditions to such large $p_T$ that there is at most only one such hard charm production process in a heavy-ion collision. Since in this case the two leading $D$~mesons are produced in the same hard process of two colliding nucleons of the nuclei, the reference curve ``without QGP'' corresponds simply to p+p collisions
and therefore can also be measured.

If we increase the trigger of the leading (subleading) $D$~meson to $p_{T;1}^D>25 \, {\rm GeV}$ ($p_{T;2}^D>15 \, {\rm GeV}$), most of these $D$~meson pairs (more than 75\,\%) are from charm quarks that have been produced in the same hard process. Thus, to save computation time and enhance statistics, we assume in the following that this is true for all of these high-energy $D$~mesons, which makes it unnecessary to also simulate the soft charm quarks in each event.

\begin{figure}
\includegraphics[width=1.0\linewidth]{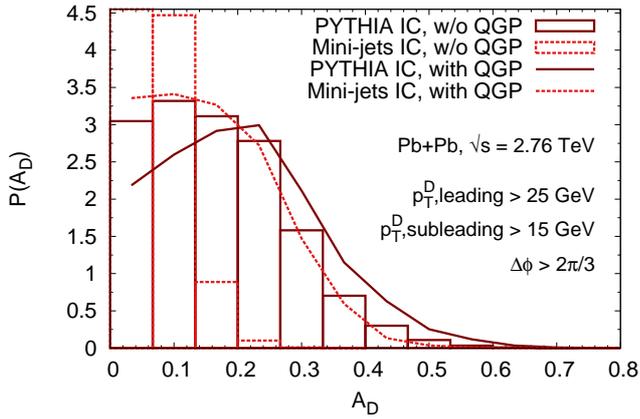}
\caption{(Color online) Distribution of the $D$~meson momentum imbalance~$A_D$ at LHC. The initial heavy quarks are obtained with either PYTHIA or the mini-jet model. The triggers are $p_{T;1}^D>25 \, {\rm GeV}$, $p_{T;2}^D>15 \, {\rm GeV}$, and $\Delta\phi>2\pi/3$.
As a note, the height of the smallest $A_D$ bin for mini-jet initial conditions without QGP (value of 9.5) is larger than the plot range.
}
\label{fig:ad_25_15_23pi}
\end{figure}

Figure~\ref{fig:ad_25_15_23pi} shows the momentum imbalance of $D$~mesons with the higher triggers for two different initial charm quark conditions.
In the mini-jet model charm quarks are produced back-to-back with the same initial $p_T$, leading to an $A_D$ of zero. The finite $A_D$ distribution in Fig.~\ref{fig:ad_25_15_23pi} without the QGP only stems  from the fragmentation of these charm quarks to $D$~mesons, which results in a small momentum imbalance. With initial conditions from PYTHIA also initial and final state radiation during the charm production process is present. Thus, the initial $p_T$ of the two produced charm quarks are not the same any more, which already leads to a strong momentum imbalance.

In both scenarios the $A_D$ distribution is shifted to higher values if interactions with the QGP are allowed, although the shift is more pronounced for mini-jet initial conditions than for distributions from PYTHIA. The small shift with PYTHIA initial conditions might be too small to be measureable.
The reason for this small shift lies in the strong initial momentum imbalance and the following two nearly balancing effects. If the leading $D$~meson loses some energy, $A_D$ decreases; if the subleading $D$~meson loses some energy, $A_D$ increases. In contrast, if the initial imbalance is small like in the mini-jet scenario, $A_D$ increases no matter which one of the $D$~mesons loses energy.
Since for mini-jet initial conditions no fluctuations in the initial $p_T$ of the two charm quarks are present, the observed momentum imbalance indeed comes from the anticipated picture of different path lengths through the medium. One of the high-energy charm quarks leaves the medium with nearly no interactions and the other loses some amount of its energy, but the corresponding $D$~meson still passes the trigger. 

Since (nearly) all $D$ meson pairs stem from the same hard event due to the high triggers, the curves without and with QGP correspond to p+p and A+A collisions, respectively. 
Although the shift in $A_D$ is small for PYTHIA initial conditions, it is comparable to what has been measured for the jet momentum imbalance~$A_J$ in p+p and A+A collisions \cite{Aad:2010bu,*Chatrchyan:2011sx,*Chatrchyan:2012nia}. Thus, we expect that the difference should also be measurable for $D$~mesons for these high triggers.

The shift in $A_D$ for PYTHIA initial conditions is considerably smaller than in Fig.~\ref{fig:ad_10_4_23pi} for two reasons: 
1) a strong charm energy loss results in $D$ mesons not passing the high triggers,
2) the low triggers in Fig.~\ref{fig:ad_10_4_23pi} allow also that $D$ mesons that are not from the same process pass the trigger, which affects the momentum imbalance with and without QGP differently, resulting in a larger difference between the two curves.

To quantify the effect of different path lengths on $A_D$, we define the path length imbalance \cite{Senzel:2013dta}
\begin{equation}
 L_D = \frac{L_{1}^D - L_{2}^D}{L_{1}^D + L_{2}^D} \ ,
\end{equation}
where $L_{1}^D$ ($L_{2}^D$) is the length of the path the leading (subleading) $D$~meson traversed in the QGP.

\begin{figure}
\includegraphics[width=1.0\linewidth]{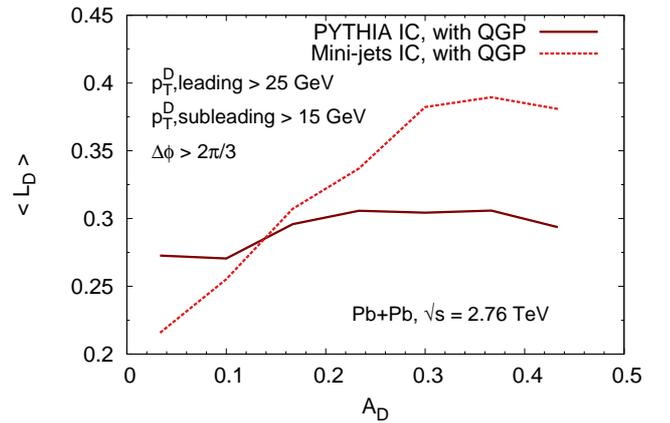}
\caption{(Color online) Mean path length imbalance $L_D$ of $D$~meson pairs as a function of their momentum imbalance $A_D$ at LHC for initial charm quarks obtained with PYTHIA and the mini-jet model. The same trigger conditions as in Fig.~\ref{fig:ad_25_15_23pi} are applied.}
\label{fig:ad_li_25_15_23pi}
\end{figure}

In Fig.~\ref{fig:ad_li_25_15_23pi} the mean path length imbalance $L_D$ of $D$~meson pairs is shown as a function of their momentum imbalance $A_D$ in heavy-ion collisions at LHC.
For mini-jet initial conditions a strong correlation between momentum and path length imbalance is visible. Thus, the momentum imbalance can be explained by different path lengths of the two corresponding charm quarks. Such a strong correlation is only present since the initial momentum imbalance within the mini-jet model is very small (cf.\ Fig.~\ref{fig:ad_25_15_23pi}).

In contrast, PYTHIA initial conditions already feature a strong momentum imbalance. Consequently, the correlation between momentum and path length imbalance is only modest. Most of the momentum imbalance stems from fluctuations in the initial charm pair distribution.

\section{Conclusions}
We proposed as a new observable the transverse momentum imbalance $A_D$ of $D$ mesons. Due to the tagged nature of charm flavored particles, this observable offers insights on the interaction of charm quarks with other constituents of the QGP. Within the partonic transport model BAMPS we simulated charm quarks in heavy ion collisions at the LHC and calculated the $A_D$ distribution for two sets of triggers. With low triggers a strong momentum imbalance $A_D$ is observed, which should be clearly measurable. However, most of the two leading $D$ mesons do not stem from the same hard process. Consequently, we increased the triggers such that most of the two leading $D$ mesons are created in the same hard process and observed a smaller difference in $A_D$ with and without QGP, which should be still measurable in the mini-jet scenario, but more difficult with PYTHIA initial conditions. We argue that in a realistic scenario with  initial charm quarks obtained with PYTHIA most of the observed momentum imbalance comes from momentum fluctuations in the initial charm pair distribution and only a minor fraction can be attributed to different path lengths in the QGP. Within the leading order mini-jet model it is the other way around. No initial momentum fluctuations are present and the observed momentum imbalance can be explained solely by a path length imbalance of the two charm quarks.

It would be highly interesting to compare the momentum imbalance $A_D$ of $D$ mesons to experimental data. Furthermore, it would be intriguing to include also radiative processes  \cite{Abir:2011jb,Fochler:2013epa} for charm quarks and investigate their effects on $A_D$. 
Comparisons between $A_D$ and $A_J$ distributions of heavy flavor tagged jets would also be an interesting future study.

\section*{Acknowledgements}
This work was supported by the Bundesministerium f\"ur Bildung und Forschung (BMBF), the NSFC under grant No.\ 11275103, HGS-HIRe and the Helmholtz International Center for FAIR within the framework of the LOEWE program launched by the State of Hesse. Numerical computations have been performed at the Center for Scientific Computing (CSC).

\bibliography{hq,text}

\begin{thebibliography}{35}%
\makeatletter
\providecommand \@ifxundefined [1]{%
 \@ifx{#1\undefined}
}%
\providecommand \@ifnum [1]{%
 \ifnum #1\expandafter \@firstoftwo
 \else \expandafter \@secondoftwo
 \fi
}%
\providecommand \@ifx [1]{%
 \ifx #1\expandafter \@firstoftwo
 \else \expandafter \@secondoftwo
 \fi
}%
\providecommand \natexlab [1]{#1}%
\providecommand \enquote  [1]{``#1''}%
\providecommand \bibnamefont  [1]{#1}%
\providecommand \bibfnamefont [1]{#1}%
\providecommand \citenamefont [1]{#1}%
\providecommand \href@noop [0]{\@secondoftwo}%
\providecommand \href [0]{\begingroup \@sanitize@url \@href}%
\providecommand \@href[1]{\@@startlink{#1}\@@href}%
\providecommand \@@href[1]{\endgroup#1\@@endlink}%
\providecommand \@sanitize@url [0]{\catcode `\\12\catcode `\$12\catcode
  `\&12\catcode `\#12\catcode `\^12\catcode `\_12\catcode `\%12\relax}%
\providecommand \@@startlink[1]{}%
\providecommand \@@endlink[0]{}%
\providecommand \url  [0]{\begingroup\@sanitize@url \@url }%
\providecommand \@url [1]{\endgroup\@href {#1}{\urlprefix }}%
\providecommand \urlprefix  [0]{URL }%
\providecommand \Eprint [0]{\href }%
\providecommand \doibase [0]{http://dx.doi.org/}%
\providecommand \selectlanguage [0]{\@gobble}%
\providecommand \bibinfo  [0]{\@secondoftwo}%
\providecommand \bibfield  [0]{\@secondoftwo}%
\providecommand \translation [1]{[#1]}%
\providecommand \BibitemOpen [0]{}%
\providecommand \bibitemStop [0]{}%
\providecommand \bibitemNoStop [0]{.\EOS\space}%
\providecommand \EOS [0]{\spacefactor3000\relax}%
\providecommand \BibitemShut  [1]{\csname bibitem#1\endcsname}%
\let\auto@bib@innerbib\@empty
\bibitem [{\citenamefont {M\"uller}\ \emph {et~al.}(2012)\citenamefont
  {M\"uller}, \citenamefont {Schukraft},\ and\ \citenamefont
  {Wyslouch}}]{Muller:2012zq}%
  \BibitemOpen
  \bibfield  {author} {\bibinfo {author} {\bibfnamefont {B.}~\bibnamefont
  {M\"uller}}, \bibinfo {author} {\bibfnamefont {J.}~\bibnamefont {Schukraft}},
  \ and\ \bibinfo {author} {\bibfnamefont {B.}~\bibnamefont {Wyslouch}},\
  }\href {\doibase 10.1146/annurev-nucl-102711-094910} {\bibfield  {journal}
  {\bibinfo  {journal} {Ann.Rev.Nucl.Part.Sci.}\ }\textbf {\bibinfo {volume}
  {62}},\ \bibinfo {pages} {361} (\bibinfo {year} {2012})},\ \Eprint
  {http://arxiv.org/abs/1202.3233} {arXiv:1202.3233 [hep-ex]} \BibitemShut
  {NoStop}%
\bibitem [{\citenamefont {Aad}\ \emph {et~al.}(2010)\citenamefont {Aad} \emph
  {et~al.}}]{Aad:2010bu}%
  \BibitemOpen
  \bibfield  {author} {\bibinfo {author} {\bibfnamefont {G.}~\bibnamefont
  {Aad}} \emph {et~al.} (\bibinfo {collaboration} {ATLAS Collaboration}),\
  }\href {\doibase 10.1103/PhysRevLett.105.252303} {\bibfield  {journal}
  {\bibinfo  {journal} {Phys.Rev.Lett.}\ }\textbf {\bibinfo {volume} {105}},\
  \bibinfo {pages} {252303} (\bibinfo {year} {2010})},\ \Eprint
  {http://arxiv.org/abs/1011.6182} {arXiv:1011.6182 [hep-ex]} \BibitemShut
  {NoStop}%
\bibitem [{\citenamefont {Chatrchyan}\ \emph {et~al.}(2011)\citenamefont
  {Chatrchyan} \emph {et~al.}}]{Chatrchyan:2011sx}%
  \BibitemOpen
  \bibfield  {author} {\bibinfo {author} {\bibfnamefont {S.}~\bibnamefont
  {Chatrchyan}} \emph {et~al.} (\bibinfo {collaboration} {CMS Collaboration}),\
  }\href {\doibase 10.1103/PhysRevC.84.024906} {\bibfield  {journal} {\bibinfo
  {journal} {Phys.Rev.}\ }\textbf {\bibinfo {volume} {C84}},\ \bibinfo {pages}
  {024906} (\bibinfo {year} {2011})},\ \Eprint {http://arxiv.org/abs/1102.1957}
  {arXiv:1102.1957 [nucl-ex]} \BibitemShut {NoStop}%
\bibitem [{\citenamefont {Chatrchyan}\ \emph {et~al.}(2012)\citenamefont
  {Chatrchyan} \emph {et~al.}}]{Chatrchyan:2012nia}%
  \BibitemOpen
  \bibfield  {author} {\bibinfo {author} {\bibfnamefont {S.}~\bibnamefont
  {Chatrchyan}} \emph {et~al.} (\bibinfo {collaboration} {CMS Collaboration}),\
  }\href {\doibase 10.1016/j.physletb.2012.04.058} {\bibfield  {journal}
  {\bibinfo  {journal} {Phys.Lett.}\ }\textbf {\bibinfo {volume} {B712}},\
  \bibinfo {pages} {176} (\bibinfo {year} {2012})},\ \Eprint
  {http://arxiv.org/abs/1202.5022} {arXiv:1202.5022 [nucl-ex]} \BibitemShut
  {NoStop}%
\bibitem [{\citenamefont {Qin}\ and\ \citenamefont
  {Muller}(2011)}]{Qin:2010mn}%
  \BibitemOpen
  \bibfield  {author} {\bibinfo {author} {\bibfnamefont {G.-Y.}\ \bibnamefont
  {Qin}}\ and\ \bibinfo {author} {\bibfnamefont {B.}~\bibnamefont {Muller}},\
  }\href {\doibase 10.1103/PhysRevLett.108.189904,
  10.1103/PhysRevLett.106.162302} {\bibfield  {journal} {\bibinfo  {journal}
  {Phys.Rev.Lett.}\ }\textbf {\bibinfo {volume} {106}},\ \bibinfo {pages}
  {162302} (\bibinfo {year} {2011})},\ \Eprint {http://arxiv.org/abs/1012.5280}
  {arXiv:1012.5280 [hep-ph]} \BibitemShut {NoStop}%
\bibitem [{\citenamefont {Casalderrey-Solana}\ \emph
  {et~al.}(2011)\citenamefont {Casalderrey-Solana}, \citenamefont {Milhano},\
  and\ \citenamefont {Wiedemann}}]{CasalderreySolana:2011rq}%
  \BibitemOpen
  \bibfield  {author} {\bibinfo {author} {\bibfnamefont {J.}~\bibnamefont
  {Casalderrey-Solana}}, \bibinfo {author} {\bibfnamefont {J.}~\bibnamefont
  {Milhano}}, \ and\ \bibinfo {author} {\bibfnamefont {U.}~\bibnamefont
  {Wiedemann}},\ }\href {\doibase 10.1088/0954-3899/38/12/124086} {\bibfield
  {journal} {\bibinfo  {journal} {J.Phys.}\ }\textbf {\bibinfo {volume}
  {G38}},\ \bibinfo {pages} {124086} (\bibinfo {year} {2011})},\ \Eprint
  {http://arxiv.org/abs/1107.1964} {arXiv:1107.1964 [hep-ph]} \BibitemShut
  {NoStop}%
\bibitem [{\citenamefont {He}\ \emph {et~al.}(2012)\citenamefont {He},
  \citenamefont {Vitev},\ and\ \citenamefont {Zhang}}]{He:2011pd}%
  \BibitemOpen
  \bibfield  {author} {\bibinfo {author} {\bibfnamefont {Y.}~\bibnamefont
  {He}}, \bibinfo {author} {\bibfnamefont {I.}~\bibnamefont {Vitev}}, \ and\
  \bibinfo {author} {\bibfnamefont {B.-W.}\ \bibnamefont {Zhang}},\ }\href
  {\doibase 10.1016/j.physletb.2012.05.054} {\bibfield  {journal} {\bibinfo
  {journal} {Phys.Lett.}\ }\textbf {\bibinfo {volume} {B713}},\ \bibinfo
  {pages} {224} (\bibinfo {year} {2012})},\ \Eprint
  {http://arxiv.org/abs/1105.2566} {arXiv:1105.2566 [hep-ph]} \BibitemShut
  {NoStop}%
\bibitem [{\citenamefont {Young}\ \emph {et~al.}(2011)\citenamefont {Young},
  \citenamefont {Schenke}, \citenamefont {Jeon},\ and\ \citenamefont
  {Gale}}]{Young:2011qx}%
  \BibitemOpen
  \bibfield  {author} {\bibinfo {author} {\bibfnamefont {C.}~\bibnamefont
  {Young}}, \bibinfo {author} {\bibfnamefont {B.}~\bibnamefont {Schenke}},
  \bibinfo {author} {\bibfnamefont {S.}~\bibnamefont {Jeon}}, \ and\ \bibinfo
  {author} {\bibfnamefont {C.}~\bibnamefont {Gale}},\ }\href {\doibase
  10.1103/PhysRevC.84.024907} {\bibfield  {journal} {\bibinfo  {journal}
  {Phys.Rev.}\ }\textbf {\bibinfo {volume} {C84}},\ \bibinfo {pages} {024907}
  (\bibinfo {year} {2011})},\ \Eprint {http://arxiv.org/abs/1103.5769}
  {arXiv:1103.5769 [nucl-th]} \BibitemShut {NoStop}%
\bibitem [{\citenamefont {Renk}(2012)}]{Renk:2012cx}%
  \BibitemOpen
  \bibfield  {author} {\bibinfo {author} {\bibfnamefont {T.}~\bibnamefont
  {Renk}},\ }\href {\doibase 10.1103/PhysRevC.85.064908} {\bibfield  {journal}
  {\bibinfo  {journal} {Phys.Rev.}\ }\textbf {\bibinfo {volume} {C85}},\
  \bibinfo {pages} {064908} (\bibinfo {year} {2012})},\ \Eprint
  {http://arxiv.org/abs/1202.4579} {arXiv:1202.4579 [hep-ph]} \BibitemShut
  {NoStop}%
\bibitem [{\citenamefont {Coleman-Smith}\ and\ \citenamefont
  {M\"uller}(2012)}]{ColemanSmith:2012vr}%
  \BibitemOpen
  \bibfield  {author} {\bibinfo {author} {\bibfnamefont {C.~E.}\ \bibnamefont
  {Coleman-Smith}}\ and\ \bibinfo {author} {\bibfnamefont {B.}~\bibnamefont
  {M\"uller}},\ }\href {\doibase 10.1103/PhysRevC.86.054901} {\bibfield
  {journal} {\bibinfo  {journal} {Phys.Rev.}\ }\textbf {\bibinfo {volume}
  {C86}},\ \bibinfo {pages} {054901} (\bibinfo {year} {2012})},\ \Eprint
  {http://arxiv.org/abs/1205.6781} {arXiv:1205.6781 [hep-ph]} \BibitemShut
  {NoStop}%
\bibitem [{\citenamefont {Ma}(2013)}]{Ma:2013pha}%
  \BibitemOpen
  \bibfield  {author} {\bibinfo {author} {\bibfnamefont {G.-L.}\ \bibnamefont
  {Ma}},\ }\href {\doibase 10.1103/PhysRevC.87.064901} {\bibfield  {journal}
  {\bibinfo  {journal} {Phys.Rev.}\ }\textbf {\bibinfo {volume} {C87}},\
  \bibinfo {pages} {064901} (\bibinfo {year} {2013})},\ \Eprint
  {http://arxiv.org/abs/1304.2841} {arXiv:1304.2841 [nucl-th]} \BibitemShut
  {NoStop}%
\bibitem [{\citenamefont {Senzel}\ \emph {et~al.}(2013)\citenamefont {Senzel},
  \citenamefont {Fochler}, \citenamefont {Uphoff}, \citenamefont {Xu},\ and\
  \citenamefont {Greiner}}]{Senzel:2013dta}%
  \BibitemOpen
  \bibfield  {author} {\bibinfo {author} {\bibfnamefont {F.}~\bibnamefont
  {Senzel}}, \bibinfo {author} {\bibfnamefont {O.}~\bibnamefont {Fochler}},
  \bibinfo {author} {\bibfnamefont {J.}~\bibnamefont {Uphoff}}, \bibinfo
  {author} {\bibfnamefont {Z.}~\bibnamefont {Xu}}, \ and\ \bibinfo {author}
  {\bibfnamefont {C.}~\bibnamefont {Greiner}},\ }\href@noop {} {\  (\bibinfo
  {year} {2013})},\ \Eprint {http://arxiv.org/abs/1309.1657} {arXiv:1309.1657
  [hep-ph]} \BibitemShut {NoStop}%
\bibitem [{\citenamefont {Uphoff}\ \emph {et~al.}(2010)\citenamefont {Uphoff},
  \citenamefont {Fochler}, \citenamefont {Xu},\ and\ \citenamefont
  {Greiner}}]{Uphoff:2010sh}%
  \BibitemOpen
  \bibfield  {author} {\bibinfo {author} {\bibfnamefont {J.}~\bibnamefont
  {Uphoff}}, \bibinfo {author} {\bibfnamefont {O.}~\bibnamefont {Fochler}},
  \bibinfo {author} {\bibfnamefont {Z.}~\bibnamefont {Xu}}, \ and\ \bibinfo
  {author} {\bibfnamefont {C.}~\bibnamefont {Greiner}},\ }\href {\doibase
  10.1103/PhysRevC.82.044906} {\bibfield  {journal} {\bibinfo  {journal}
  {Phys.Rev.}\ }\textbf {\bibinfo {volume} {C82}},\ \bibinfo {pages} {044906}
  (\bibinfo {year} {2010})},\ \Eprint {http://arxiv.org/abs/1003.4200}
  {arXiv:1003.4200 [hep-ph]} \BibitemShut {NoStop}%
\bibitem [{\citenamefont {Colamaria}(2013)}]{Colamaria:2013una}%
  \BibitemOpen
  \bibfield  {author} {\bibinfo {author} {\bibfnamefont {F.}~\bibnamefont
  {Colamaria}} (\bibinfo {collaboration} {ALICE collaboration}),\ }\href@noop
  {} {\  (\bibinfo {year} {2013})},\ \Eprint {http://arxiv.org/abs/1310.3621}
  {arXiv:1310.3621 [hep-ex]} \BibitemShut {NoStop}%
\bibitem [{\citenamefont {Thomas}(2013)}]{DeepaThomasfortheALICE:2013oga}%
  \BibitemOpen
  \bibfield  {author} {\bibinfo {author} {\bibfnamefont {D.}~\bibnamefont
  {Thomas}} (\bibinfo {collaboration} {ALICE Collaboration}),\ }\href@noop {}
  {\  (\bibinfo {year} {2013})},\ \Eprint {http://arxiv.org/abs/1308.5464}
  {arXiv:1308.5464 [hep-ex]} \BibitemShut {NoStop}%
\bibitem [{\citenamefont {Zhu}\ \emph {et~al.}(2007)\citenamefont {Zhu} \emph
  {et~al.}}]{Zhu:2006er}%
  \BibitemOpen
  \bibfield  {author} {\bibinfo {author} {\bibfnamefont {X.}~\bibnamefont
  {Zhu}} \emph {et~al.},\ }\href {\doibase 10.1016/j.physletb.2007.01.072}
  {\bibfield  {journal} {\bibinfo  {journal} {Phys.Lett.}\ }\textbf {\bibinfo
  {volume} {B647}},\ \bibinfo {pages} {366} (\bibinfo {year} {2007})},\ \Eprint
  {http://arxiv.org/abs/hep-ph/0604178} {arXiv:hep-ph/0604178} \BibitemShut
  {NoStop}%
\bibitem [{\citenamefont {Zhu}\ \emph {et~al.}(2008)\citenamefont {Zhu},
  \citenamefont {Xu},\ and\ \citenamefont {Zhuang}}]{Zhu:2007ne}%
  \BibitemOpen
  \bibfield  {author} {\bibinfo {author} {\bibfnamefont {X.}~\bibnamefont
  {Zhu}}, \bibinfo {author} {\bibfnamefont {N.}~\bibnamefont {Xu}}, \ and\
  \bibinfo {author} {\bibfnamefont {P.}~\bibnamefont {Zhuang}},\ }\href
  {\doibase 10.1103/PhysRevLett.100.152301} {\bibfield  {journal} {\bibinfo
  {journal} {Phys.Rev.Lett.}\ }\textbf {\bibinfo {volume} {100}},\ \bibinfo
  {pages} {152301} (\bibinfo {year} {2008})},\ \Eprint
  {http://arxiv.org/abs/0709.0157} {arXiv:0709.0157 [nucl-th]} \BibitemShut
  {NoStop}%
\bibitem [{\citenamefont {Gossiaux}\ \emph {et~al.}(2009)\citenamefont
  {Gossiaux}, \citenamefont {Bierkandt},\ and\ \citenamefont
  {Aichelin}}]{Gossiaux:2009mk}%
  \BibitemOpen
  \bibfield  {author} {\bibinfo {author} {\bibfnamefont {P.~B.}\ \bibnamefont
  {Gossiaux}}, \bibinfo {author} {\bibfnamefont {R.}~\bibnamefont {Bierkandt}},
  \ and\ \bibinfo {author} {\bibfnamefont {J.}~\bibnamefont {Aichelin}},\
  }\href {\doibase 10.1103/PhysRevC.79.044906} {\bibfield  {journal} {\bibinfo
  {journal} {Phys.Rev.}\ }\textbf {\bibinfo {volume} {C79}},\ \bibinfo {pages}
  {044906} (\bibinfo {year} {2009})},\ \Eprint {http://arxiv.org/abs/0901.0946}
  {arXiv:0901.0946 [hep-ph]} \BibitemShut {NoStop}%
\bibitem [{\citenamefont {Younus}\ and\ \citenamefont
  {Srivastava}(2013)}]{Younus:2013be}%
  \BibitemOpen
  \bibfield  {author} {\bibinfo {author} {\bibfnamefont {M.}~\bibnamefont
  {Younus}}\ and\ \bibinfo {author} {\bibfnamefont {D.~K.}\ \bibnamefont
  {Srivastava}},\ }\href {\doibase 10.1088/0954-3899/40/6/065004} {\bibfield
  {journal} {\bibinfo  {journal} {J.Phys.}\ }\textbf {\bibinfo {volume}
  {G40}},\ \bibinfo {pages} {065004} (\bibinfo {year} {2013})},\ \Eprint
  {http://arxiv.org/abs/1301.5762} {arXiv:1301.5762 [nucl-th]} \BibitemShut
  {NoStop}%
\bibitem [{\citenamefont {Nahrgang}\ \emph {et~al.}(2013)\citenamefont
  {Nahrgang}, \citenamefont {Aichelin}, \citenamefont {Gossiaux},\ and\
  \citenamefont {Werner}}]{Nahrgang:2013saa}%
  \BibitemOpen
  \bibfield  {author} {\bibinfo {author} {\bibfnamefont {M.}~\bibnamefont
  {Nahrgang}}, \bibinfo {author} {\bibfnamefont {J.}~\bibnamefont {Aichelin}},
  \bibinfo {author} {\bibfnamefont {P.~B.}\ \bibnamefont {Gossiaux}}, \ and\
  \bibinfo {author} {\bibfnamefont {K.}~\bibnamefont {Werner}},\ }\href@noop {}
  {\  (\bibinfo {year} {2013})},\ \Eprint {http://arxiv.org/abs/1305.3823}
  {arXiv:1305.3823 [hep-ph]} \BibitemShut {NoStop}%
\bibitem [{\citenamefont {Xu}\ and\ \citenamefont {Greiner}(2005)}]{Xu:2004mz}%
  \BibitemOpen
  \bibfield  {author} {\bibinfo {author} {\bibfnamefont {Z.}~\bibnamefont
  {Xu}}\ and\ \bibinfo {author} {\bibfnamefont {C.}~\bibnamefont {Greiner}},\
  }\href {\doibase 10.1103/PhysRevC.71.064901} {\bibfield  {journal} {\bibinfo
  {journal} {Phys.Rev.}\ }\textbf {\bibinfo {volume} {C71}},\ \bibinfo {pages}
  {064901} (\bibinfo {year} {2005})},\ \Eprint
  {http://arxiv.org/abs/hep-ph/0406278} {arXiv:hep-ph/0406278} \BibitemShut
  {NoStop}%
\bibitem [{\citenamefont {Xu}\ and\ \citenamefont {Greiner}(2007)}]{Xu:2007aa}%
  \BibitemOpen
  \bibfield  {author} {\bibinfo {author} {\bibfnamefont {Z.}~\bibnamefont
  {Xu}}\ and\ \bibinfo {author} {\bibfnamefont {C.}~\bibnamefont {Greiner}},\
  }\href {\doibase 10.1103/PhysRevC.76.024911} {\bibfield  {journal} {\bibinfo
  {journal} {Phys.Rev.}\ }\textbf {\bibinfo {volume} {C76}},\ \bibinfo {pages}
  {024911} (\bibinfo {year} {2007})},\ \Eprint
  {http://arxiv.org/abs/hep-ph/0703233} {arXiv:hep-ph/0703233} \BibitemShut
  {NoStop}%
\bibitem [{\citenamefont {Sjostrand}\ \emph {et~al.}(2006)\citenamefont
  {Sjostrand}, \citenamefont {Mrenna},\ and\ \citenamefont
  {Skands}}]{Sjostrand:2006za}%
  \BibitemOpen
  \bibfield  {author} {\bibinfo {author} {\bibfnamefont {T.}~\bibnamefont
  {Sjostrand}}, \bibinfo {author} {\bibfnamefont {S.}~\bibnamefont {Mrenna}}, \
  and\ \bibinfo {author} {\bibfnamefont {P.}~\bibnamefont {Skands}},\
  }\href@noop {} {\bibfield  {journal} {\bibinfo  {journal} {JHEP}\ }\textbf
  {\bibinfo {volume} {05}},\ \bibinfo {pages} {026} (\bibinfo {year} {2006})},\
  \Eprint {http://arxiv.org/abs/hep-ph/0603175} {arXiv:hep-ph/0603175}
  \BibitemShut {NoStop}%
\bibitem [{\citenamefont {Gunion}\ and\ \citenamefont
  {Bertsch}(1982)}]{Gunion:1981qs}%
  \BibitemOpen
  \bibfield  {author} {\bibinfo {author} {\bibfnamefont {J.}~\bibnamefont
  {Gunion}}\ and\ \bibinfo {author} {\bibfnamefont {G.}~\bibnamefont
  {Bertsch}},\ }\href {\doibase 10.1103/PhysRevD.25.746} {\bibfield  {journal}
  {\bibinfo  {journal} {Phys.Rev.}\ }\textbf {\bibinfo {volume} {D25}},\
  \bibinfo {pages} {746} (\bibinfo {year} {1982})}\BibitemShut {NoStop}%
\bibitem [{\citenamefont {Fochler}\ \emph {et~al.}(2013)\citenamefont
  {Fochler}, \citenamefont {Uphoff}, \citenamefont {Xu},\ and\ \citenamefont
  {Greiner}}]{Fochler:2013epa}%
  \BibitemOpen
  \bibfield  {author} {\bibinfo {author} {\bibfnamefont {O.}~\bibnamefont
  {Fochler}}, \bibinfo {author} {\bibfnamefont {J.}~\bibnamefont {Uphoff}},
  \bibinfo {author} {\bibfnamefont {Z.}~\bibnamefont {Xu}}, \ and\ \bibinfo
  {author} {\bibfnamefont {C.}~\bibnamefont {Greiner}},\ }\href {\doibase
  10.1103/PhysRevD.88.014018} {\bibfield  {journal} {\bibinfo  {journal}
  {Phys.Rev.}\ }\textbf {\bibinfo {volume} {D88}},\ \bibinfo {pages} {014018}
  (\bibinfo {year} {2013})},\ \Eprint {http://arxiv.org/abs/1302.5250}
  {arXiv:1302.5250 [hep-ph]} \BibitemShut {NoStop}%
\bibitem [{\citenamefont {Uphoff}\ \emph {et~al.}(2014)\citenamefont {Uphoff},
  \citenamefont {Fochler}, \citenamefont {Senzel}, \citenamefont {Wesp},
  \citenamefont {Xu} \emph {et~al.}}]{Uphoff:2014cba}%
  \BibitemOpen
  \bibfield  {author} {\bibinfo {author} {\bibfnamefont {J.}~\bibnamefont
  {Uphoff}}, \bibinfo {author} {\bibfnamefont {O.}~\bibnamefont {Fochler}},
  \bibinfo {author} {\bibfnamefont {F.}~\bibnamefont {Senzel}}, \bibinfo
  {author} {\bibfnamefont {C.}~\bibnamefont {Wesp}}, \bibinfo {author}
  {\bibfnamefont {Z.}~\bibnamefont {Xu}},  \emph {et~al.},\ }\href@noop {} {\
  (\bibinfo {year} {2014})},\ \Eprint {http://arxiv.org/abs/1401.1364}
  {arXiv:1401.1364 [hep-ph]} \BibitemShut {NoStop}%
\bibitem [{\citenamefont {Peterson}\ \emph {et~al.}(1983)\citenamefont
  {Peterson}, \citenamefont {Schlatter}, \citenamefont {Schmitt},\ and\
  \citenamefont {Zerwas}}]{Peterson:1982ak}%
  \BibitemOpen
  \bibfield  {author} {\bibinfo {author} {\bibfnamefont {C.}~\bibnamefont
  {Peterson}}, \bibinfo {author} {\bibfnamefont {D.}~\bibnamefont {Schlatter}},
  \bibinfo {author} {\bibfnamefont {I.}~\bibnamefont {Schmitt}}, \ and\
  \bibinfo {author} {\bibfnamefont {P.~M.}\ \bibnamefont {Zerwas}},\ }\href
  {\doibase 10.1103/PhysRevD.27.105} {\bibfield  {journal} {\bibinfo  {journal}
  {Phys.Rev.}\ }\textbf {\bibinfo {volume} {D27}},\ \bibinfo {pages} {105}
  (\bibinfo {year} {1983})}\BibitemShut {NoStop}%
\bibitem [{\citenamefont {Uphoff}\ \emph {et~al.}(2011)\citenamefont {Uphoff},
  \citenamefont {Fochler}, \citenamefont {Xu},\ and\ \citenamefont
  {Greiner}}]{Uphoff:2011ad}%
  \BibitemOpen
  \bibfield  {author} {\bibinfo {author} {\bibfnamefont {J.}~\bibnamefont
  {Uphoff}}, \bibinfo {author} {\bibfnamefont {O.}~\bibnamefont {Fochler}},
  \bibinfo {author} {\bibfnamefont {Z.}~\bibnamefont {Xu}}, \ and\ \bibinfo
  {author} {\bibfnamefont {C.}~\bibnamefont {Greiner}},\ }\href {\doibase
  10.1103/PhysRevC.84.024908} {\bibfield  {journal} {\bibinfo  {journal}
  {Phys.Rev.}\ }\textbf {\bibinfo {volume} {C84}},\ \bibinfo {pages} {024908}
  (\bibinfo {year} {2011})},\ \Eprint {http://arxiv.org/abs/1104.2295}
  {arXiv:1104.2295 [hep-ph]} \BibitemShut {NoStop}%
\bibitem [{\citenamefont {Uphoff}\ \emph {et~al.}(2012)\citenamefont {Uphoff},
  \citenamefont {Fochler}, \citenamefont {Xu},\ and\ \citenamefont
  {Greiner}}]{Uphoff:2012gb}%
  \BibitemOpen
  \bibfield  {author} {\bibinfo {author} {\bibfnamefont {J.}~\bibnamefont
  {Uphoff}}, \bibinfo {author} {\bibfnamefont {O.}~\bibnamefont {Fochler}},
  \bibinfo {author} {\bibfnamefont {Z.}~\bibnamefont {Xu}}, \ and\ \bibinfo
  {author} {\bibfnamefont {C.}~\bibnamefont {Greiner}},\ }\href {\doibase
  10.1016/j.physletb.2012.09.069} {\bibfield  {journal} {\bibinfo  {journal}
  {Phys.Lett.}\ }\textbf {\bibinfo {volume} {B717}},\ \bibinfo {pages} {430}
  (\bibinfo {year} {2012})},\ \Eprint {http://arxiv.org/abs/1205.4945}
  {arXiv:1205.4945 [hep-ph]} \BibitemShut {NoStop}%
\bibitem [{\citenamefont {Conesa~del Valle}(2013)}]{delValle:2012qw}%
  \BibitemOpen
  \bibfield  {author} {\bibinfo {author} {\bibfnamefont {Z.}~\bibnamefont
  {Conesa~del Valle}} (\bibinfo {collaboration} {ALICE Collaboration}),\ }\href
  {\doibase 10.1016/j.nuclphysa.2013.01.060} {\bibfield  {journal} {\bibinfo
  {journal} {Nucl.Phys.}\ }\textbf {\bibinfo {volume} {A904-905}},\ \bibinfo
  {pages} {178c} (\bibinfo {year} {2013})},\ \Eprint
  {http://arxiv.org/abs/1212.0385} {arXiv:1212.0385 [nucl-ex]} \BibitemShut
  {NoStop}%
\bibitem [{\citenamefont {Sakai}(2013)}]{Sakai:2013ata}%
  \BibitemOpen
  \bibfield  {author} {\bibinfo {author} {\bibfnamefont {S.}~\bibnamefont
  {Sakai}} (\bibinfo {collaboration} {ALICE Collaboration}),\ }\href {\doibase
  10.1016/j.nuclphysa.2013.02.102} {\bibfield  {journal} {\bibinfo  {journal}
  {Nucl.Phys.}\ }\textbf {\bibinfo {volume} {A904-905}},\ \bibinfo {pages}
  {661c} (\bibinfo {year} {2013})}\BibitemShut {NoStop}%
\bibitem [{\citenamefont {Abelev}\ \emph {et~al.}(2013)\citenamefont {Abelev}
  \emph {et~al.}}]{Abelev:2013lca}%
  \BibitemOpen
  \bibfield  {author} {\bibinfo {author} {\bibfnamefont {B.}~\bibnamefont
  {Abelev}} \emph {et~al.} (\bibinfo {collaboration} {ALICE Collaboration}),\
  }\href@noop {} {\  (\bibinfo {year} {2013})},\ \Eprint
  {http://arxiv.org/abs/1305.2707} {arXiv:1305.2707 [nucl-ex]} \BibitemShut
  {NoStop}%
\bibitem [{\citenamefont {Abelev}\ \emph {et~al.}(2012)\citenamefont {Abelev}
  \emph {et~al.}}]{ALICE:2012ab}%
  \BibitemOpen
  \bibfield  {author} {\bibinfo {author} {\bibfnamefont {B.}~\bibnamefont
  {Abelev}} \emph {et~al.} (\bibinfo {collaboration} {ALICE Collaboration}),\
  }\href {\doibase 10.1007/JHEP09(2012)112} {\bibfield  {journal} {\bibinfo
  {journal} {JHEP}\ }\textbf {\bibinfo {volume} {1209}},\ \bibinfo {pages}
  {112} (\bibinfo {year} {2012})},\ \Eprint {http://arxiv.org/abs/1203.2160}
  {arXiv:1203.2160 [nucl-ex]} \BibitemShut {NoStop}%
\bibitem [{\citenamefont {Grelli}(2013)}]{Grelli:2012yv}%
  \BibitemOpen
  \bibfield  {author} {\bibinfo {author} {\bibfnamefont {A.}~\bibnamefont
  {Grelli}} (\bibinfo {collaboration} {ALICE Collaboration}),\ }\href {\doibase
  10.1016/j.nuclphysa.2013.02.096} {\bibfield  {journal} {\bibinfo  {journal}
  {Nucl.Phys.}\ }\textbf {\bibinfo {volume} {A904-905}},\ \bibinfo {pages}
  {635c} (\bibinfo {year} {2013})},\ \Eprint {http://arxiv.org/abs/1210.7332}
  {arXiv:1210.7332 [hep-ex]} \BibitemShut {NoStop}%
\bibitem [{\citenamefont {Abir}\ \emph {et~al.}(2012)\citenamefont {Abir},
  \citenamefont {Greiner}, \citenamefont {Martinez}, \citenamefont {Mustafa},\
  and\ \citenamefont {Uphoff}}]{Abir:2011jb}%
  \BibitemOpen
  \bibfield  {author} {\bibinfo {author} {\bibfnamefont {R.}~\bibnamefont
  {Abir}}, \bibinfo {author} {\bibfnamefont {C.}~\bibnamefont {Greiner}},
  \bibinfo {author} {\bibfnamefont {M.}~\bibnamefont {Martinez}}, \bibinfo
  {author} {\bibfnamefont {M.~G.}\ \bibnamefont {Mustafa}}, \ and\ \bibinfo
  {author} {\bibfnamefont {J.}~\bibnamefont {Uphoff}},\ }\href {\doibase
  10.1103/PhysRevD.85.054012} {\bibfield  {journal} {\bibinfo  {journal}
  {Phys.Rev.}\ }\textbf {\bibinfo {volume} {D85}},\ \bibinfo {pages} {054012}
  (\bibinfo {year} {2012})},\ \Eprint {http://arxiv.org/abs/1109.5539}
  {arXiv:1109.5539 [hep-ph]} \BibitemShut {NoStop}%
\end{thebibliography}%

\end{document}